\documentclass[
 reprint,
 superscriptaddress,
 amsmath,amssymb,
 aip,jcp
]{revtex4-2}

\usepackage{graphicx}

\usepackage{dcolumn}
\usepackage{bm}
\usepackage{hyperref}

\renewcommand{\vec}{\textbf}

\usepackage[dvipsnames]{xcolor}
\usepackage{soul}

\begin{document}

\title{Synthetic data enable experiments in atomistic machine learning}

\author{John L. A. Gardner}
\author{Zo\'e{} Faure Beaulieu}
\author{Volker L. Deringer}
\email{volker.deringer@chem.ox.ac.uk}
\affiliation{Department of Chemistry, Inorganic Chemistry Laboratory, University of Oxford, Oxford OX1 3QR, UK}

\begin{abstract}
Machine-learning models are increasingly used to predict properties of atoms in chemical systems.
There have been major advances in developing descriptors and regression frameworks for this task, typically starting from (relatively) small sets of quantum-mechanical reference data.
Larger datasets of this kind are becoming available, but remain expensive to generate.
Here we demonstrate the use of a large dataset that we have ``synthetically'' labelled with per-atom energies from an existing ML potential model.
The cheapness of this process, compared to the quantum-mechanical ground truth, allows us to generate millions of datapoints, in turn enabling rapid experimentation with atomistic ML models from the small- to the large-data regime.
This approach allows us here to compare regression frameworks in depth, and to explore visualisation based on learned representations.
We also show that learning synthetic data labels can be a useful pre-training task for subsequent fine-tuning on small datasets.
In the future, we expect that our open-sourced dataset, and similar ones, will be useful in rapidly exploring deep-learning models in the limit of abundant chemical data.
\end{abstract}

\maketitle

\section*{Introduction}

Chemical research aims to understand existing, and to discover new, molecules and materials. 
The vast size of compositional and configurational chemical space means that physical experiments will quickly reach their limits for these tasks.\cite{reymondExploringChemicalSpace2012, polishchukEstimationSizeDruglike2013, restrepoChemicalSpaceLimits2022} 
Digital ``experiments'', powered by large datasets and machine learning (ML) models, provide high-throughput approaches to chemical discovery, 
and can help to answer questions that their physical counterparts on their own can not.\cite{curtaroloHighthroughputHighwayComputational2013, coleyAutonomousDiscoveryChemical2020, coleyAutonomousDiscoveryChemical2020a, kauweCanMachineLearning2020}
However, because ML methods generally rely on large datasets rather than on empirical physical knowledge, they require new insight into the methodology itself -- one example in this context is the active research into interpretability and explainability of ML models. \cite{dybowskiInterpretableMachineLearning2020, oviedoInterpretableExplainableMachine2022}

Among the central tasks in ML for chemistry is the prediction of atomistic properties as a function of a given atom's chemical environment. Atomistic ML models have now been developed to predict scalar ({\em e.g.}, isotropic chemical shifts),\cite{paruzzoChemicalShiftsMolecular2018, chakerNMRShiftsAluminosilicate2019a} vector ({\em e.g.}, dipole moments),\cite{veitPredictingMolecularDipole2020a} and higher-order tensor properties ({\em e.g.}, the dielectric response).\cite{grisafiSymmetryAdaptedMachineLearning2018} 
ML methods are also increasingly enabling accurate, large-scale atomistic simulations based on the ``learning'' of a given quantum-mechanical potential-energy surface. 
Widely used approaches for ML interatomic potential models include neural networks (NNs), \cite{behlerGeneralizedNeuralNetworkRepresentation2007,schuttSchNetContinuousfilterConvolutional2017a,gasteigerDirectionalMessagePassing2022,huForceNetGraphNeural2021a} 
kernel-based methods,\cite{bartokGaussianApproximationPotentials2010, chmielaMachineLearningAccurate2017}
and linear fitting.\cite{thompsonSpectralNeighborAnalysis2015, shapeevMomentTensorPotentials2016b}
The most suitable choice out of these options may depend on the task and chemical domain. \cite{pinheiroChoosingRightMolecular2021}
For instance, the ability of NNs to scalably learn compressed, hierarchical, and meaningful representations has allowed them to converge to ``chemical accuracy'' in the small-molecule setting on the established QM9 dataset.\cite{ramakrishnanQuantumChemistryStructures2014, Lubbers2018, Schuett2018, Unke2019}

When exploring a new chemical system for which there is no established, large dataset, it is not necessarily obvious which model class will be suitable for a given task, or how a model will perform.
Unfortunately, creating the high-quality, quantum-mechanically accurate data needed to train such ML models is very expensive.
For instance, using density-functional theory (DFT) to generate and label the 1.2 million structures within the OC20 database required the use of large-scale compute resources, and millions of CPU hours.\cite{chanussotOpenCatalyst20202021}
This cost often limits the size of dataset available when exploring different model classes on new chemical domains, favouring simpler models with high data economy over more complex ones that benefit from large data quantities.

\begin{figure}
    \centering
    \includegraphics[width=\linewidth]{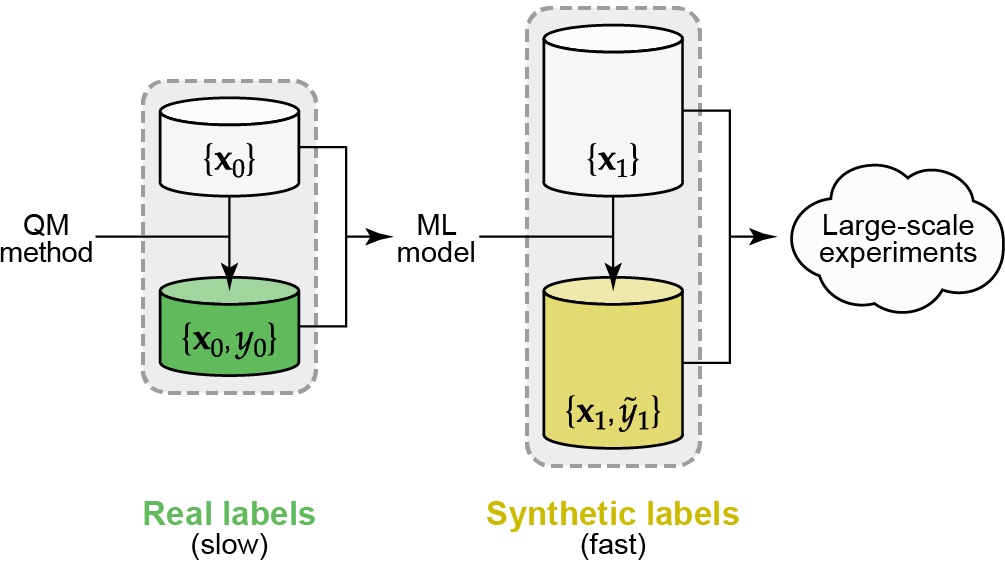}
    \caption{Synthetic data for atomistic ML. {\em Left:} Quantum-mechanical (QM) data are used to label a set of structures, $\vec{x}_{0}$, with energy and force data, $y_{0}$, and these serve as input for an ML model of the potential-energy surface. {\em Right:} QM labels are expensive, and so we here use an existing ML model to cheaply generate and label a much larger dataset. The data in this set are ``synthetic'' as they are not labelled with the ground-truth QM method itself, yet represent its behaviour.
    (Note that whilst the QM method describes energies and forces on atoms, our synthetic dataset is labelled only with per-atom energies in the present study.)
    }
    \label{fig:synthetic_data}
\end{figure}

\begin{figure*}
    \centering
    \includegraphics[width=\textwidth]{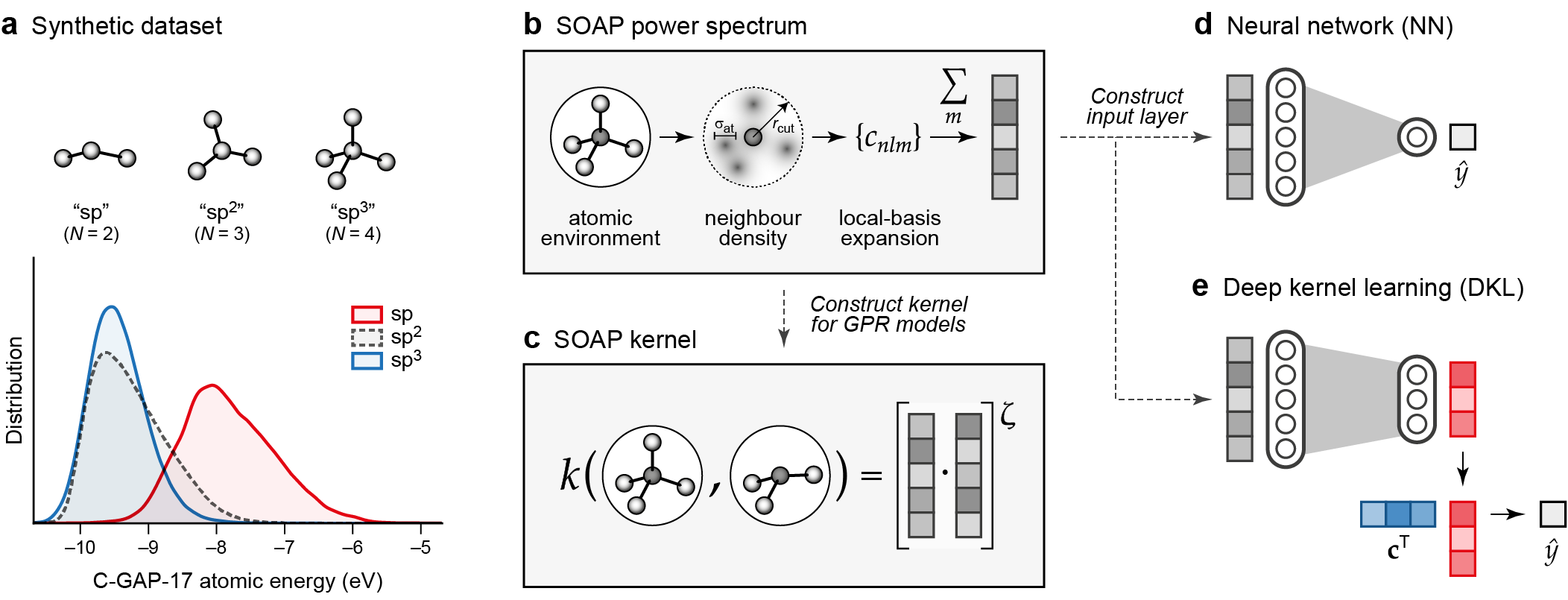}
    \caption{Overview of ML approaches used in the present work.
    (a) A synthetic dataset of atomic energies, predicted by the C-GAP-17 model, \cite{deringerMachineLearningBased2017c} for different categories of carbon environments as sketched. The distributions are shown by kernel density estimates, normalised to the same value for each (note there are much fewer sp than sp$^{2}$ atoms overall). Energies are referenced to that of a free atom.
    (b) Smooth Overlap of Atomic Positions (SOAP) power spectrum.\cite{bartokRepresentingChemicalEnvironments2013a} The power-spectrum vector is an invariant fingerprint of the atomic environment, and illustrated in grey.
    (c) Construction of the SOAP kernel, as a dot-product of the power-spectrum vectors for two atomic environments, raised to a power of $\zeta$.
    (d) Neural-network model. We use the power-spectrum vector [{\em cf.} panel (b)] to directly construct the input layer, and train a network to predict a new value, $\hat{y}$, from this.
    (e) Deep kernel learning. A neural network is used to learn a compressed representation of atomic environments, indicated in red, from the original SOAP vectors. Gaussian process regression is then used to make predictions in this compressed space, using a learned set of coefficients, $\vec{c}$, and the similarity of a new point to each entry in the data set.
    Panel (b) is adapted from Ref.\ \citenum{deringerGaussianProcessRegression2021}, originally published under a CC BY licence (https://creativecommons.org/licenses/by/4.0/).
    }
    \label{fig:methods}
\end{figure*}

Here we demonstrate the use of synthetic data labels, obtained from an existing ML potential model (Fig.\ \ref{fig:synthetic_data}), as a means to sidestep the high computational cost of quantum-accurate labelling that would otherwise be required for experimenting with atomistic ML approaches.
Concretely, we introduce an open-sourced dataset containing 22.9 million atomic environments drawn from ML-driven molecular-dynamics (MD) simulations of diverse disordered carbon structures, subsequently labelled in less than a day on local, consumer-level compute. 
The size of this dataset enables us to study the behaviour of different ML models in the small- and large-data limits.

\section*{Dataset}

Our dataset consists of 546 independent MD trajectories describing the melt-quenching and thermal annealing of elemental carbon. The development of ML potentials for carbon \cite{khaliullinGraphitediamondPhaseCoexistence2010, deringerMachineLearningBased2017c,roweAccurateTransferableMachine2020a,willmanQuantumAccurateSNAP2020b,shaiduSystematicApproachGenerating2021} and their application to scientific problems \cite{thiemannDefectDependentCorrugationGraphene2021, karasuluAcceleratingPredictionLarge2022, golzeAccurateComputationalPrediction2022} have been widely documented in the literature, and the existence of established potentials such as C-GAP-17 (Ref.\ \citenum{deringerMachineLearningBased2017c}) means that there is a direct route for creating synthetic data.

Initial randomised configurations of 200 atoms per cell at varying densities, from 1.0 to 3.5\,g\,cm$^{-3}$ in 0.1\,g\,cm$^{-3}$ increments, were generated using ASE.\cite{larsenAtomicSimulationEnvironment2017}
Each structure then underwent an MD simulation driven by C-GAP-17, as implemented in LAMMPS.\cite{thompsonLAMMPSFlexibleSimulation2022} First, each structure was melted at 9000\,K for 5\,ps before being quenched to 300\,K over a further 5\,ps. Second, each structure was reheated to a specific temperature at which it was annealed for 100\,ps, before finally being cooled back down to 300\,K over 50\,ps. The annealing temperatures ranged from 2000 to 4000\,K in 100\,K increments. These protocols are in line with prior quenching-and-annealing type simulations with empirical and machine-learned potentials. \cite{powlesSelfassemblySpBonded2009, detomasStructuralPredictionGraphitization2017, deringerAtomisticUnderstandingDisordered2018, wangStructurePoreSize2022}

The resulting database captures a wide variety of chemical environments, including graphitic structures, buckyball-esque clusters, grains of cubic and hexagonal diamond, and tetrahedral amorphous carbon. 
Every atom in the dataset was labelled using the C-GAP-17 potential, which predicts per-atom energies as a function of a given atom's environment. \cite{deringerMachineLearningBased2017c} Figure \ref{fig:methods}a shows the distribution of these energies in the dataset, categorised in a simplified manner by their coordination number: ``sp'' as in carbon chains ($N = 2$), ``sp$^{2}$'' as in graphite ($N = 3$), and ``sp$^{3}$'' as in diamond ($N = 4$). The energies for the sp$^{2}$ and sp$^{3}$ environments are rather similar, consistent with the very similar energy of graphite and diamond; those for sp atoms are notably higher.

\section*{Methods}

{\bf Structural descriptors.} We describe (``featurise'') atomic environments using the Smooth Overlap of Atomic Positions (SOAP) technique.\cite{bartokRepresentingChemicalEnvironments2013a} SOAP is based on the idea of a local-basis expansion of the atomic neighbour density and subsequent construction of a rotationally invariant power spectrum (Fig.\ \ref{fig:methods}b).\cite{bartokRepresentingChemicalEnvironments2013a} 
Initially developed as a similarity measure between pairs of local neighbourhood densities, SOAP can also provide a descriptor of a single local environment, and be used as input to other ML techniques. \cite{kocerNovelApproachDescribe2019a, karamadOrbitalGraphConvolutional2020a, xiaPredictionMaterialProperties2021a} 
In the present work, we use SOAP power-spectrum vectors in two ways: to construct kernel matrices for Gaussian process regression (GPR), and as a base from which to learn richer and compressed descriptions using neural network models (Fig.\ \ref{fig:methods}c--e).

The SOAP descriptor is controlled by four (hyper-) parameters.\cite{bartokRepresentingChemicalEnvironments2013a} Two convergence parameters, $n_{\max}$ and $l_{\max}$, control the number of radial and angular basis functions, respectively; the radial cut-off, $r_{\textrm{cut}}$, defines the locality of the environment, and a Gaussian broadening width, $\sigma_{\textrm{at}}$, controls the smoothness of the atomic neighbourhood densities.
Here, descriptor vectors pre-calculated using $(n_{\max}, l_{\max}) = (12, 6)$ led to convergence for the average value in the SOAP similarity matrix for a 200-atom structure to within 0.01\%, as compared to $(16, 16)$.
Values of $3.7$\,\AA{} and $0.5$\,\AA{} for $r_{\textrm{cut}}$ and $\sigma_{\textrm{at}}$, respectively, were used, in line with the settings for the C-GAP-17 model.\cite{deringerMachineLearningBased2017c} 

{\bf Gaussian process regression (GPR).}
GPR non-parametrically fits a probabilistic model to high-dimensional data. 
For a detailed introduction to GPR, see Ref.\ \citenum{rasmussenGaussianProcessesMachine2006}, and for its applications in chemistry, see Ref.\ \citenum{deringerGaussianProcessRegression2021}.
At a high level, prediction at a test point, $\mathbf{x}^\prime$, involves calculating its similarity to each data location in the training set, $\mathbf{x}_{i}$, using a specified kernel, $k$.
Each of these similarities, $k(\mathbf{x}_i, \mathbf{x}^\prime)$, then modulates a coefficient, $c_i$, learned during training, such that the prediction is
\begin{equation}
\hat{y}(\mathbf{x}^\prime) = \sum_i^N c_i \cdot k(\mathbf{x}_i, \mathbf{x}^\prime) \quad \equiv \mathbf{c} \cdot k(\mathbf{X}, \mathbf{x}').
\end{equation}
In this work, we evaluated $k(\mathbf{x}, \mathbf{x}^\prime)$ as the dot product of the respective SOAP power-spectrum vectors, raised to the power of $\zeta = 4$ as is common practice.\cite{deringerGaussianProcessRegression2021}

For an exact implementation of GPR, the time complexity for predicting $\hat{y}(\mathbf{x}^\prime)$ is $O(N)$, where $N$ is the number of training example pairs, $\{\mathbf{x}_i, y_i\}$. However, solving for $\mathbf{c}$ during training entails an $O(N^3)$ time and $O(N^2)$ storage cost. In practical terms, this limits ``full GPR'' to at most a few thousand data points.\cite{rasmussenGaussianProcessesMachine2006}
One approach to circumventing this unfavourable scaling is referred to as sparse GPR,\cite{deringerGaussianProcessRegression2021} 
which only considers $M$ representative data locations when making predictions.
Prediction time complexity is therefore $O(M)$, while training entails $O(M^2N + M^3)$ time and $O(NM + M^2)$ space scaling. Provided $M \ll N$, this can significantly increase the amount of data that can be used for training in practice. 
In the present work, we used $M=5000$, and varied $N$ up to $10^{6}$.

{\bf Neural-network (NN) models.}
Artificial NNs can provably represent any function given sufficient parameterisation. \cite{barronUniversalApproximationBounds1993a, lecunDeepLearning2015} 
For an overview of the inspiration for, workings of, and theory behind NNs, we refer to Ref.\ \citenum{schmidhuberDeepLearningNeural2015a}. In brief, NNs make predictions by repeatedly applying alternating linear and non-linear transforms, parameterised by weights and biases. These are learned using backpropagation to iteratively reduce a loss function.

Throughout this work, we train NNs using standard forward and backward propagation techniques using the Adam optimiser \cite{kingmaAdamMethodStochastic2017} and CELU activation functions, \cite{barronContinuouslyDifferentiableExponential2017b} all as implemented in PyTorch.\cite{paszkeAutomaticDifferentiationPyTorch2017}
The performance of a deep NN depends heavily on the choice of hyperparameters for the model architecture and training, including the depth and width of the network, and the learning rate of the optimiser.
We establish optimised values for these hyperparameters using an automated process: random sweep over values, and validating on a test set (see below).

{\bf Deep kernel learning (DKL).}
DKL models make predictions through the sequential application of deep NN and GPR models:\cite{wilsonDeepKernelLearning2015a, wilsonStochasticVariationalDeep2016a} the NN takes high-dimensional data as input and outputs a compressed representation in a space where the Euclidean distance between two data points, relative to the learned length scale of the GPR model, is representative of their (dis-) similarity.

During training, the parameters of both the NN and GPR model are jointly optimised by maximising the log posterior marginal likelihood. These models were implemented using the PyTorch and GPyTorch libraries.\cite{paszkeAutomaticDifferentiationPyTorch2017, gardnerGPyTorchBlackboxMatrixMatrix2018}

\begin{figure}
    \centering
    \includegraphics[width=\linewidth]{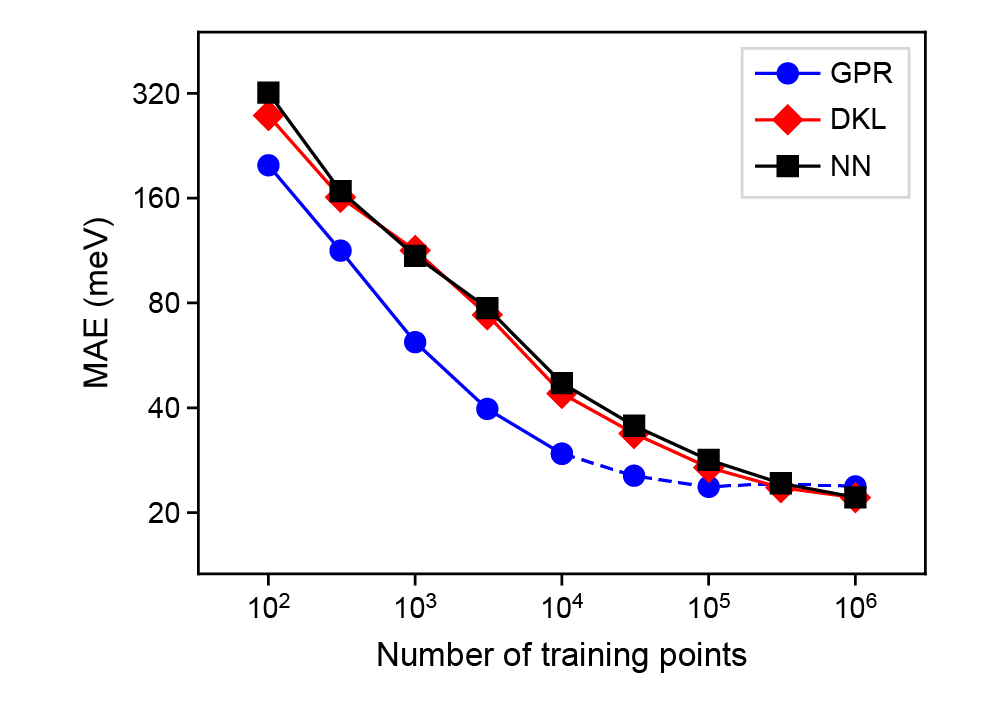}
    \caption{\label{fig:learning-curves} 
    Learning curves for our synthetic dataset.
    Mean absolute error (MAE) values for the prediction of C-GAP-17 labelled local energies, as a function of training set size (``learning curves''), for the most accurate instance of each model class.
    The dashed line indicates GPR models that required specialist compute ($>$64GB RAM) to train.
    }
\end{figure}

\section*{Learning curves}

The first result of this paper is the demonstration that our synthetic data, {\em viz.} ML atomic energies, can indeed be machine-learned, and how the quality of this learning depends on the size of the training dataset. 
In Fig.\ \ref{fig:learning-curves}, we test the ability of our GPR, NN, and DKL models to learn atomic energies from the dataset of carbon structures described above. 
We show ``learning curves'' that allow us to quantify and compare the errors of the three model classes considered. 
In each case, mean absolute error (MAE) values are quoted as averaged using a 5-fold cross-validation procedure, where the structures from a single MD trajectory are dedicated completely to either the training or the test set, to avoid training example data leakage. \cite{Morrow2022a} 
We note that this learning of ML-predicted data is related to the recently proposed ``indirect learning'' approach, \cite{Morrow2022} but it is distinctly different in that the latter does not regress per-atom energies, rather aiming to create teacher--student ML potential models.

In the low-data regime, the learning curves in Fig.\ \ref{fig:learning-curves} show the behaviour known for other atomistic ML models: the error decreases linearly on the double-logarithmic plot. There is a clear advantage of the GPR models (blue) over either network-based technique (NNs, black; DKL, red) in this regime, with $10^{4}$ data points perhaps being representative of a specialised learning problem in quantum chemistry requiring expensive data labels. However, this effect is diminished upon moving to larger datasets. Typical ML potentials use on the order of $10^{5}$ data points for training, and in this region the learning curve for the GPR models visibly saturates. We emphasise that we use sparse GPR, and so the actual number of points in the regression, $M$, is much lower than $10^{5}$; this aspect will be discussed in the following section.

Comparing the NN and DKL models side-by-side, we find no notable advantage of DKL over regular NNs in this context -- a slight gain in accuracy comes at a cost of approximately 100-fold slower prediction. In the remainder of this paper, we will therefore focus on a deeper analysis of GPR and NN models for atomistic ML, and report on numerical experiments with these two model classes.

\section*{Experiments}

\subsection*{GPR insights}

\begin{figure}
    \centering
    \includegraphics[width=\linewidth]{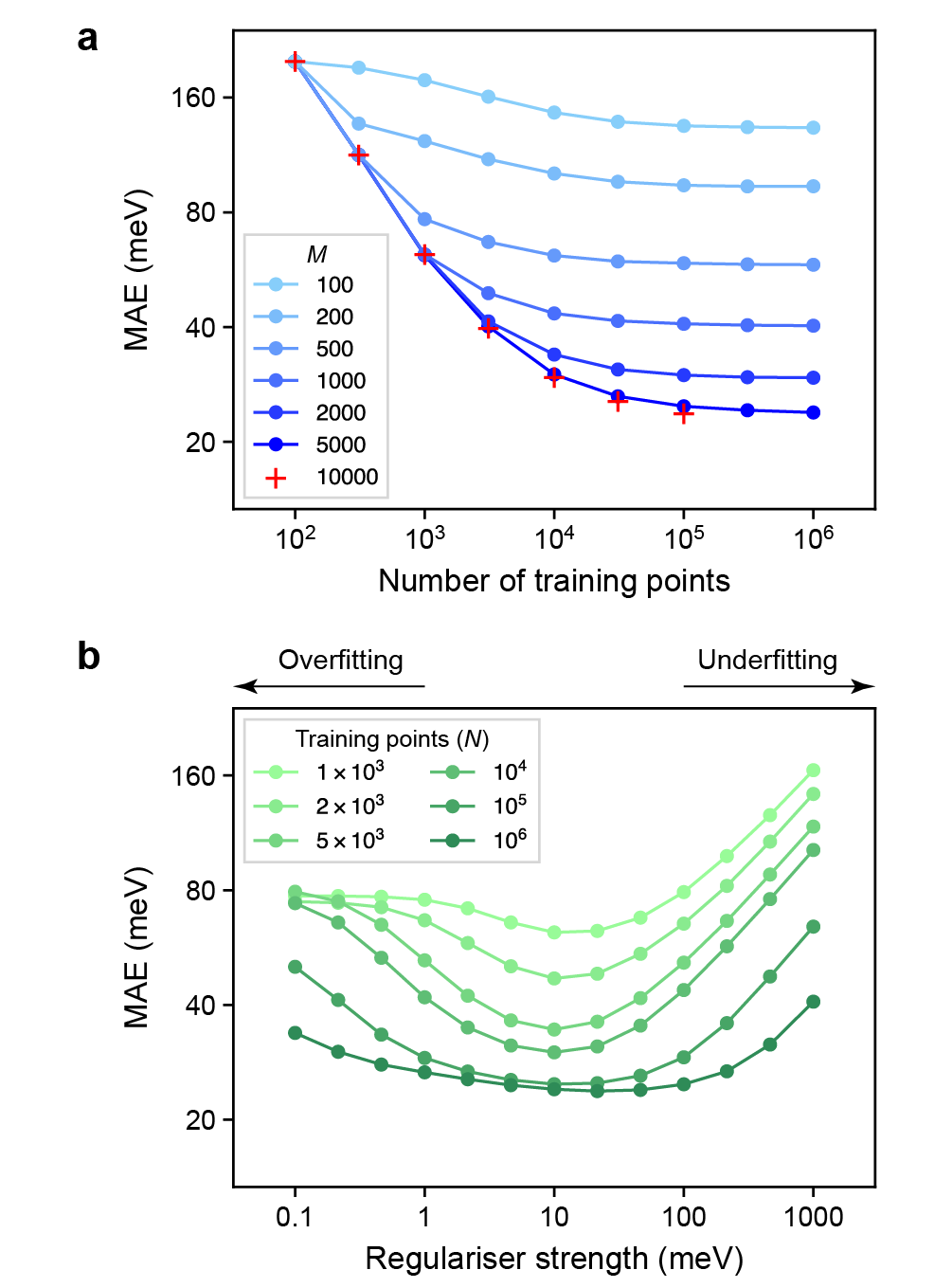}
    \caption{\label{fig:gpr-insight} Aspects of GPR models and their effect on prediction quality. 
    (a) MAE for a series of GPR models with varied numbers of representative points, $M$, and training points, $N$. We used $M=$ 5000 in the present work, limited by memory availability for $N=10^{6}$. We include results for $M=$ 10,000 as far as practicable, and find that those do not lead to substantial improvements in the region tested. 
    (b) MAE for a series of GPR models with varied regularisation terms. Too low values will cause the model to overfit to data, whereas too high values (too high ``expected error'') will diminish the quality of the prediction. The minimum value is found around 10 meV for most values of $N$, and this setting was used for all other GPR results shown in this work.
}
\end{figure}

Having identified synthetic atomic energies as a ``machine-learnable'' and readily available target quantity (Fig.\ \ref{fig:learning-curves}), we can use these synthetic data to gain further insight into GPR models.
There are two important considerations when constructing sparse GPR models that we address here. 

The first aspect is the choice of the number of representative points, $M$, that are used in the sparse GPR fit. 
In a full GPR setting, the fitting coefficients, $\mathbf{c}$, would be obtained at training time as 
\begin{equation}
    \mathbf{c} = (\mathbf{K}_{NN} + \mathbf{\Sigma})^{-1} \mathbf{y},
    \label{eq:GPR_fit_full}
\end{equation}
where $\mathbf{K}_{NN}$ is a matrix of kernel similarity values between any two data locations, {\em viz.} $(K_{NN})_{i,j} = k(\mathbf{x}_{i}, \mathbf{x}_{j})$, $\mathbf{\Sigma}$ is a regularisation term, and the vector $\mathbf{y}$ collects all labels in the training dataset. \cite{rasmussenGaussianProcessesMachine2006, deringerGaussianProcessRegression2021} In sparse GPR, as we use here, the analogous equation for obtaining the fitting coefficients reads:\cite{bartokGaussianApproximationPotentials2010}
\begin{equation}
\mathbf{c}  = [\mathbf{K}_{MM} + 
    \mathbf{K}_{MN} \mathbf{\Sigma}^{-1} \mathbf{K}_{NM}]^{-1} \mathbf{K}_{MN} \mathbf{\Sigma}^{-1} \mathbf{y},    
    \label{eq:GPR_fit_sparse}
\end{equation}
where similarly defined kernel matrices, now of different sizes, are used to quantify similarities between individual atomic environments. The resulting coefficient vector, $\mathbf{c}$, now has length $M$ (not $N$), and therefore the number of {\em representative} points, $M$, is what effectively controls the computational cost at runtime. For example, in a widely used GPR-based potential for elemental silicon, the reference database includes hundreds of thousands of atomic force components, whilst $M$ is only 9000.\cite{bartokMachineLearningGeneralPurpose2018}

In Fig.\ \ref{fig:gpr-insight}a, we show learning curves as in the previous section, but now for different values of $M$ in otherwise similar GPR models.
We find that the change from $M=$ 5000 to $M=$ 10,000 does not seem to lead to a major change any more, at least up to the range investigated.

The second aspect that we explore for our GPR models is the regularisation, controlled by the matrix $\mathbf{\Sigma}$ in Eqs.\ \ref{eq:GPR_fit_full} and \ref{eq:GPR_fit_sparse} above.
The regularisation applied during training can be interpreted as ``expected error'' of the data (in the context of interatomic potentials, this might be due to accuracy and convergence limits of the quantum-mechanical training data\cite{deringerGaussianProcessRegression2021}). 
Another interpretation is as the driving force applied during training that affects the extent to with the final fit passes through all the data points.
For simplicity, we use a constant regularisation value for all atoms in the database; that is,
\begin{equation}
    \mathbf{\Sigma} = \sigma \mathbf{I},
\end{equation}
with $\sigma = 10$ meV unless noted otherwise.
We note in passing that more adaptive approaches are possible, such as an individual regularisation for each atom, as exemplified before in GAP fitting. \cite{georgeCombiningPhononAccuracy2020}

We tested the effect of varying the regularisation value, $\sigma$, over a wide range of values -- the ease with which synthetic data labels are accessible means that we can rapidly fit many candidate models, both in terms of $\sigma$, and of the number of training points, $N$. The results are shown in Fig.\ \ref{fig:gpr-insight}b.

Interestingly, the dependence on the regularisation becomes less pronounced with larger $N$: the curves visibly flatten as one moves to larger datasets. At the same time, there is still a difference between the $N=10^{5}$ and $N=10^{6}$ curves in Fig.\ \ref{fig:gpr-insight}b, even though the learning curve itself had already ``levelled off'' (Fig.\ \ref{fig:learning-curves}).
We observe a flatter curve for the larger dataset, and a shift of the location of the minimum to higher $\sigma$, although it remains to be explored how significant the latter is. 
GPR is a fundamentally Bayesian technique, and so the behaviour seen in Fig.\ \ref{fig:gpr-insight}b can be rationalised by noting that using more data reduces the importance of any priors, in this case, of the exact value of $\sigma$.
As larger and larger datasets become used for GPR-based models, tuning of the regularisation is therefore expected to become less important.

In retrospect, both plots in Fig.\ \ref{fig:gpr-insight} seem to confirm settings that have been intuitively used in ML potential fitting using the GAP framework. \cite{deringerGaussianProcessRegression2021} We are curious whether large-scale experiments on synthetic (proxy) data can, in the future, inform the choice of regularisation and other hyperparameters in new and more complex ``real-world'' GPR models for chemistry.

\begin{figure*}
    \centering
    \includegraphics[width=\linewidth]{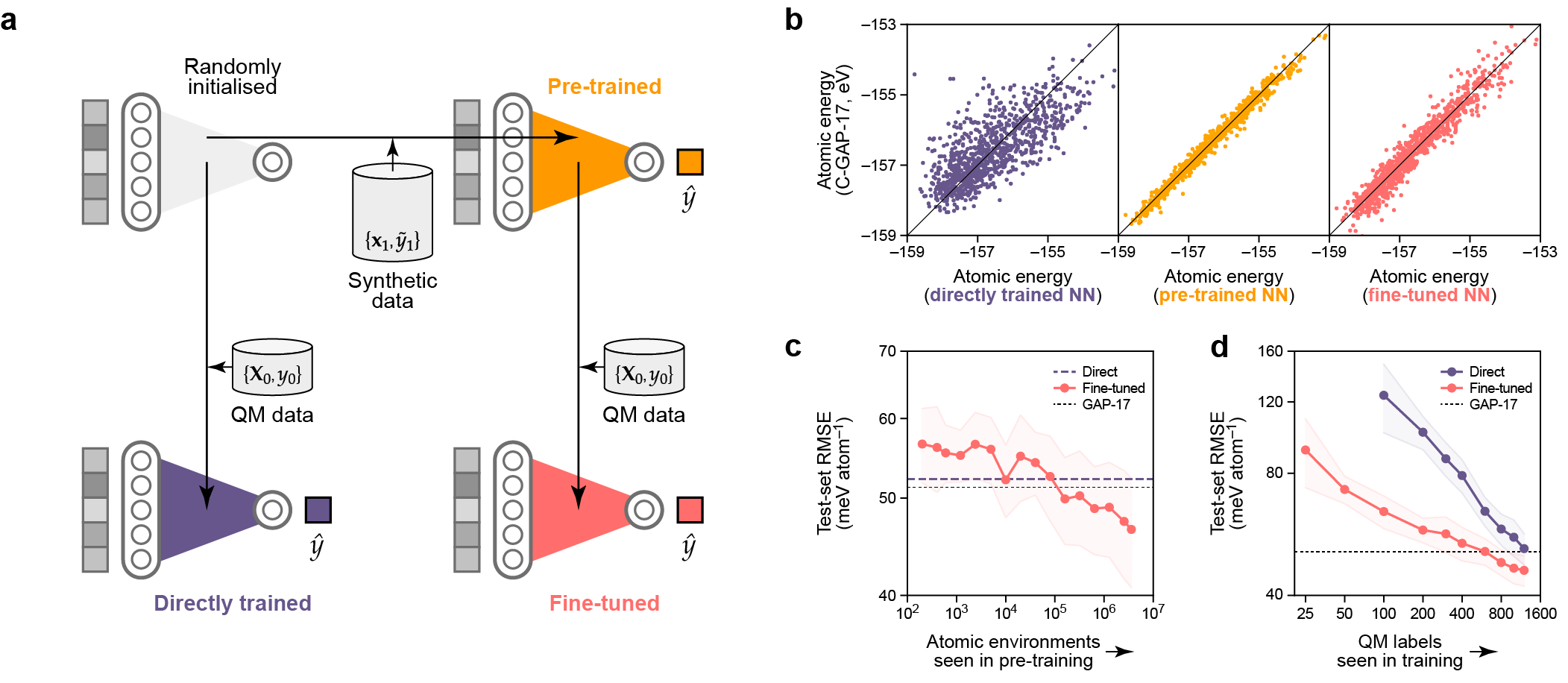}
    \caption{Pre-training neural networks with synthetic atomistic data. 
    (a) Schematic of the experiment. We pre-train an NN (orange) on local energies from a large synthetic dataset, {\bf D1} ($\equiv \left\{ \mathbf{x}_{1}, \tilde{y}_{1} \right\}$), then use the optimised parameters as a starting point for training (``fine-tuning'') another NN (red) on quantum-mechanical (``QM'') total energies from a smaller dataset, {\bf D0}.
    We compare against the more conventional approach of directly training on {\bf D0} (purple).
    (b) Parity plots for local (per-atom) energy predictions, testing against {\bf D1}, {\em i.e.}, against the performance of C-GAP-17. From left to right: the directly trained NN (which learns to assign local energies in a different manner from GAP), the pre-trained NN (with tight correlation), and finally the fine-tuned model.
    (c) Effect of varying the number of pre-training environments on final test-set accuracy when fine-tuned on the complete {\bf D0}. Low numbers of pretraining environments have little effect on the final accuracy ($\sim$ no change as compared to direct training). Increasing the number of pre-training environments beyond about $10^5$ ({\em i.e.}, roughly the number of environments in {\bf D0}) leads to increasing performance of the fine-tuned over the directly trained model.
    (d) Learning curves that show the dependence on the number of QM labels seen during training.
    To obtain a model with the same predictive power, in this case, $\sim 8 \times$ fewer QM labels are required when starting from a pre-trained NN, as compared to random initialisation.
    Note that we truncate the plot at 100 QM labels for direct training, but extend it to as low as 25 for the fine-tuned NNs.
}
    \label{fig:pre-training}
\end{figure*}

\subsection*{Pre-training}

We now move to the discussion of neural-network methodology for atomistic ML.
We hypothesised that our synthetic dataset can be used to pre-train an atomistic NN model, which can then be fine-tuned for predicting a related property.
For this approach to be useful, it needs to lead to a better final model than training an NN directly without prior information. The idea behind this experiment is sketched in Fig.\ \ref{fig:pre-training}a.

The task on which we have focused so far is to minimise the atom-wise (squared) error of our model predictions as compared to synthetic labels:
\begin{equation}
    \underset{\lambda}{\textrm{argmin}} \quad  \mathcal{L} \left( \sum_{i} \left| \tilde{\varepsilon}_i - \hat{\varepsilon}_\lambda(\mathbf{x}_i) \right|^{2} \right),
\end{equation}
where $\tilde{\varepsilon}$ are the ML atomic energies, as labelled by C-GAP-17 and used here as synthetic training data, and $\hat{\varepsilon}$ are our model's predictions of this property. The loss function, $\mathcal{L}$, is optimised with respect to the set of model parameters, $\lambda$.

The task that we ultimately want to perform is the prediction of quantum-mechanical, per-cell energies, $E$:
we have a dataset, {\bf D0}, consisting of pairs $\{\mathbf{X}, E_{\mathrm{DFT}}\}$, where $\mathbf{X}$ is the set of descriptor vectors, $\mathbf{x}_{i}$, together describing all atoms in a given unit cell, and $E_{\mathrm{DFT}}$ is the per-cell energy as calculated using DFT (in this proof-of-concept, {\bf D0} is the subset of all 64-atom amorphous carbon structures taken from the C-GAP-17 database \cite{deringerMachineLearningBased2017c}).
In many currently used NN models for chemistry, this task involves predicting per-cell energies as a sum of local atomic energies: \cite{behlerGeneralizedNeuralNetworkRepresentation2007, smith_ani-1_2017, Zhang2018a, Batzner2022}
\begin{equation}
\hat{E}_{c} = \sum_{i \in c} \hat{\varepsilon}_\lambda(\mathbf{x}_i).
\end{equation}
The optimisation problem then becomes
\begin{equation}
    \underset{\lambda}{\textrm{argmin}} \quad \mathcal{L} \left( \sum_{c \in \mathbf{D0}}  \left| E_{{\rm DFT}, c} - \sum_{\mathbf{x}_i \in \mathbf{X}_c} \varepsilon_\lambda(\mathbf{x}_i) \right|^{2} \right),
    \label{eqn:DFT-opt}
\end{equation}
where $E_{\mathrm{DFT}, c}$ is the ground-truth value for cell $c$ against which the model parameters $\lambda$ are optimised.

We first describe the control experiment: training a randomly initialised NN model exclusively on per-structure energies, $E_{{\rm DFT,c}}$.
The resulting model (purple in Fig. \ref{fig:pre-training}a) learns atomic energies that, when summed over a cell, predict per-cell energies with a test-set RMSE of $52$ meV atom$^{-1}$. 
This is, to within noise, the same as the original C-GAP-17 model (on its own training data!).
Interestingly, the NN model trained in this way learns to partition per-cell energies into atomic contributions in a different manner to C-GAP-17: a parity plot of these (Fig.\ \ref{fig:pre-training}b) shows only a loose correlation.
This non-uniqueness of local energies from NN models seems to be in keeping with previous findings.\cite{eckhoffMolecularFragmentsBulk2019}

Training a new NN solely on C-GAP-17 local energies for the synthetic dataset unsurprisingly leads to a model (orange in Fig. \ref{fig:pre-training}) that reproduces these quantities much more closely. Starting from this pretrained model and its set of optimised parameters, and subsequently performing the same per-cell energy optimisation procedure (Eqn. \ref{eqn:DFT-opt}), we obtain an NN (red in Fig. \ref{fig:pre-training}) that performs significantly better than direct training from a random initialisation in predicting per-cell energies, with a test-set RMSE of $46.5$ meV atom$^{-1}$ (Fig.\ \ref{fig:pre-training}d). 
The parity plots (Fig. \ref{fig:pre-training}b) show that the fine-tuned network, having been originally guided by the C-GAP-17 local energies, partitions local energies in a more similar manner to C-GAP-17 as compared to the direct training approach.

We perform preliminary tests for the role of dataset size in this pre-training procedure. 
Figure \ref{fig:pre-training}c suggests that {\bf D1} needs to be at least as large as {\bf D0} (in terms of number of atomic environments) in order for the pre-training approach to improve upon the accuracy from direct training on {\bf D0} (purple dashed line). Note that we use hyperparameters that maximise accuracy when training on the full {\bf D1} set (here, using 4 million atomic environments) for all pre-training dataset sizes investigated; thus, while Fig.\ \ref{fig:pre-training}c might seem to suggest that small amounts of pre-training are detrimental, we assume that they actually have no effect in practice.
Figure \ref{fig:pre-training}d shows that, when pre-training on the full {\bf D1} set (red), $\sim 8 \times$ fewer QM labels are required to achieve the same final accuracy compared to using the direct approach (purple). Thus we have shown initial evidence that learning to predict synthetic atomic energies can be a useful and meaningful “pre-training task” for chemistry.

We note that our pre-training can be recast as transfer learning from a lower to a higher level of quantum-aware labelling. Transfer learning for atomistic models has been demonstrated by Smith {\em et al.},\cite{smithApproachingCoupledCluster2019} although we are not aware of prior work going from ML-potential- to DFT-level accuracy, or performing this transfer learning {\em via} synthetic atomic energies (rather than per-structure energies or forces).
We also note that the pre-training of NN models is a well-documented approach in the ML literature for various applications and domains, \cite{Pennington-14-01,Krizhevsky-17-05} and that it has very recently been described in the context of interatomic potential models \cite{Zhang-22-09, Gao-22-11}
and as a means to learn general-purpose representations for atomistic structure.\cite{Gao-22-11}

\begin{figure*}
    \centering
    \includegraphics[width=\linewidth]{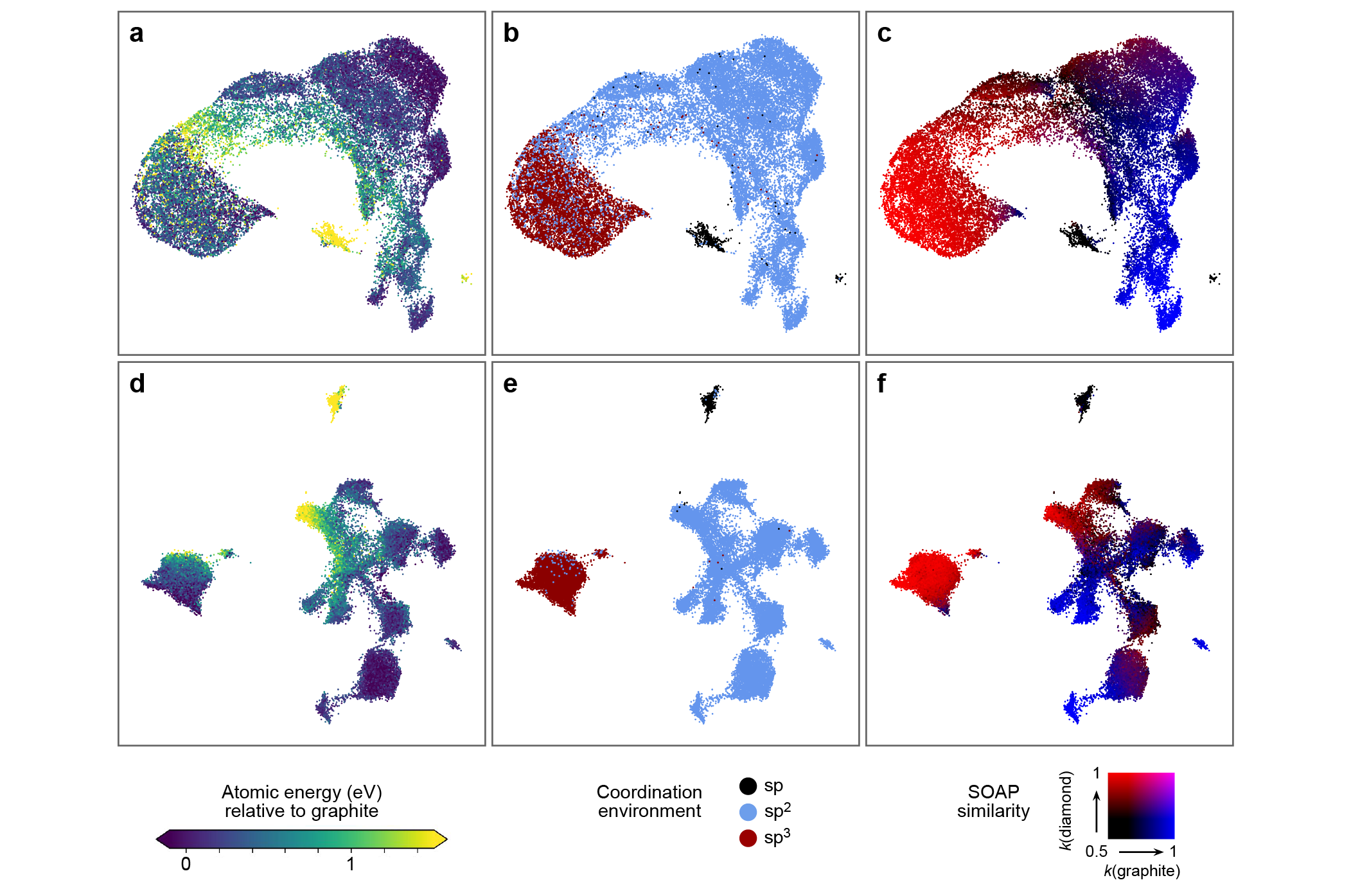}
    \caption{\label{fig:embeddings} 
    UMAP projections\cite{mcinnesUMAPUniformManifold2020a} to visualise the configurational space of the synthetic dataset, as described by the original SOAP vectors (a--c), and by the compressed representations learned by an NN trained on synthetic C-GAP-17 atomic energies (d--f).
    From left to right, we colour-code by the C-GAP-17 atomic energy relative to graphite, by coordination environment category as determined by a 1.85\,\AA{}  cut-off, and by SOAP similarity to diamond (red) and graphite (blue).
    Some clustering exists in the original SOAP space, although many strictly sp$^2$ atoms are found within the space predominantly populated by sp$^3$ carbon. At a local scale, the gradient in atomic energy is very noisy in this space.
    Compare this to the representation learned by the NN: clear clustering occurs that aligns very tightly with carbon coordination environment. Within the sp$^2$ region, further sub-clustering exists, each of which has clear meaning as highlighted by the SOAP similarity colour coding. The local gradient in atomic energy is much smoother, as is to be expected given the network has been trained on this quantity.
}
\end{figure*}

\begin{figure}
    \centering
    \includegraphics[width=\linewidth]{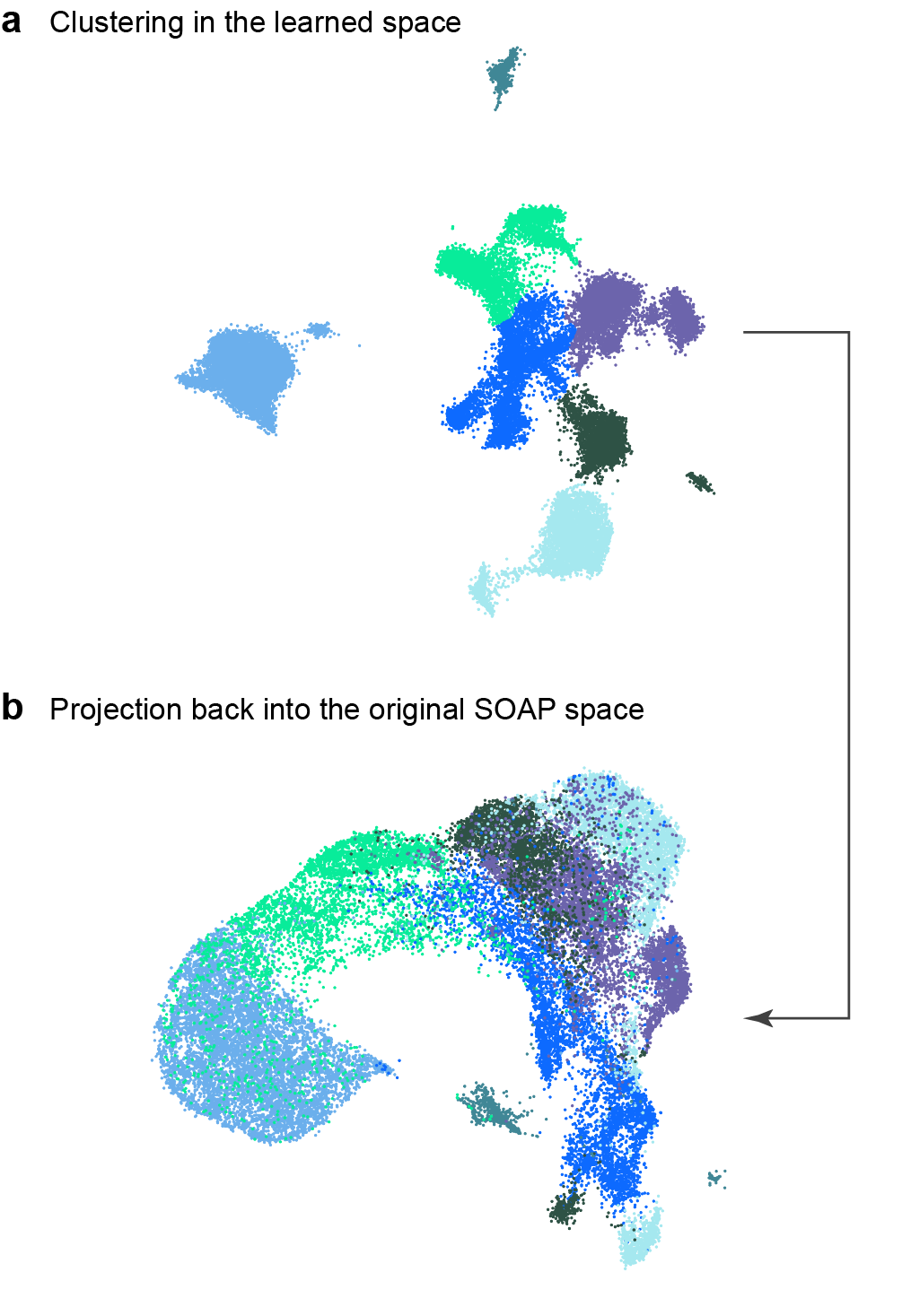}
    \caption{
    UMAP embeddings showing (a) clusters found in the NN-based map of Fig.\ \ref{fig:embeddings}d--f, and (b) projection of these cluster labels into the original SOAP space.
    Clear mixing occurs in the latter, showing that the NN is learning a different representation than the SOAP features with which it was trained, rather than some linear recombination of these.
    }
    \label{fig:maps_cluster}
\end{figure}

\subsection*{Embedding and visualisation}

We finally illustrate the usefulness of synthetic atomistic data for the visualisation of structural and chemical space. 
This is an increasingly important task in ML for chemistry, leading to what are commonly referred to as ``structure maps''.\cite{chengMappingMaterialsMolecules2020}
Out of the many recipes for creating such a map, a popular one entails (i) selecting a metric to define the (dis-) similarity, or distance, between two atomic environments, (ii) calculating this metric for each pair from a (representative) set of environments to create a distance matrix, and (iii) embedding this matrix onto a low-dimensional (2D or 3D) manifold.
A popular instantiation of this recipe is to use the SOAP kernel as a similarity metric, and to use a non-linear dimensionality reduction technique, such as UMAP or t-SNE, to embed the distance matrix. \cite{deComparingMoleculesSolids2016, chengMappingMaterialsMolecules2020} 
Such SOAP-based maps have been reported for pristine and chemically functionalised forms of carbon\cite{caroReactivityAmorphousCarbon2018, shiresVisualizingEnergyLandscapes2021, golzeAccurateComputationalPrediction2022} and can help in understanding local environments -- {\em e.g.}, in assigning the chemical character of a given atom beyond the simplified ``sp'' / ``sp$^{2}$'' / ``sp$^{3}$'' labels. \cite{caroReactivityAmorphousCarbon2018}  

Our present work explores the ability of an NN model to generate analogous 2D maps of chemical structure. Outputs from hidden layers of an atomistic NN can be used to visualise structural space.\cite{westermayrNeuralNetworksKernel2020} We investigate such visualisation for a model that has been trained on the synthetic dataset introduced above, and draw comparison with earlier work on carbon. \cite{caroReactivityAmorphousCarbon2018}
We adapt the above recipe by using the Euclidean distance between NN penultimate hidden-layer representations as a dissimilarity metric, and embed the resulting matrix using UMAP,\cite{mcinnesUMAPUniformManifold2020a} a general approach not limited to chemistry. \cite{Dorkenwald-22-11}

Figure \ref{fig:embeddings} shows the resulting structure maps. 
We use 30,000 atomic environments selected at random from the dataset presented above.
The upper row (Fig.\ \ref{fig:embeddings}a--c) shows a SOAP-based UMAP embedding, colour-coded by relevant properties, whereas the lower row (Fig.\ \ref{fig:embeddings}d--f) shows a UMAP embedding derived from hidden-layer representations of an NN model. The former confirms observations made before on smaller datasets: \cite{caroReactivityAmorphousCarbon2018} distinct types of environments, both energetically and structurally, can be discerned in the map by different colour coding. For example, there is a small island of structures with high local energy (yellow, Fig. \ref{fig:embeddings}a), and those correspond to the twofold-bonded ``sp'' environments, as seen from Fig.\ \ref{fig:embeddings}b. This observation is consistent with the energy distributions shown in Fig.\ \ref{fig:methods}a.   

Whilst the SOAP map does therefore capture relevant aspects, the clustering in the map produced from the learned NN representations (Fig.\ \ref{fig:embeddings}d--f) is significantly more intricate, while also aligning more strongly with our understanding (or the chemistry textbook picture) of carbon atom hybridisation. Specifically, compared to the embeddings produced from SOAP descriptors, the NN space shows a much clearer separation of sp$^2$ {\em vs} sp$^3$ atoms (dark red {\em vs} light blue in the central panels), and it also shows sub-clustering within the sp$^2$ region. We can interpret this sub-clustering by colour-coding each datapoint according to the corresponding environment's SOAP similarity to both graphite (blue) and diamond (red). These results are shown on the right-hand side of Fig. \ref{fig:embeddings}: some of the formally sp$^2$ carbons are very similar to diamond-like environments, suggesting that these are in fact dangling-bond sp$^3$ environments, further corroborated by their high local energies. This kind of structure is not made as obvious in the SOAP map in Fig.\ \ref{fig:embeddings}c.

In Fig. \ref{fig:maps_cluster}, we show further evidence that the NN has learned a different description of local structure from the original SOAP descriptors.
We performed cluster analysis in the NN-based structure map, using the BIRCH algorithm\cite{Zhang-96-06} to separate the data into 7 distinct clusters (colour-coded arbitrarily in Fig.\ \ref{fig:maps_cluster}a), and then projected the resulting cluster labels into the space of the original SOAP map (Fig.\ \ref{fig:maps_cluster}b).
Doing so shows that atoms contained within the same cluster in NN space are not necessarily co-located in SOAP space:
while the sp atoms remain isolated in the SOAP map ({\em cf.}\ Fig.\ \ref{fig:embeddings}b), the remaining clusters are clearly heavily intermixed, with some ({\em e.g.}, bright green) spanning most of the SOAP space.
Hence, the mapping between SOAP vector and NN representation is complex and highly non-linear, such that the NN is truly learning a new representation.

We therefore argue that maps based on hidden layers of atomistic NN models, trained on synthetic datasets as exemplified here, can capture aspects of both the structure and the energetics of a given material.
This is not merely a consequence of the higher flexibility of NNs -- in fact, an analogue to the SOAP-based maps shown herein would be the visualisation of the latent space of an autoencoder model. 
Instead, we here take the hidden layer following {\em supervised} learning, thereby automatically incorporating information about the data labels in the structure map (although the question how exactly this information is learned is deliberately left to the network to optimise).
There is some similarity of this approach to principal covariates regression \cite{dejongPrincipalCovariatesRegression1992} which has been combined with kernel metrics for use in atomistic ML, \cite{helfrechtStructurepropertyMapsKernel2020} and the resulting ``kernel PCovR'' was applied to silicon under pressure. \cite{deringerGaussianProcessRegression2021} Here, however, the data labels can enter into the model in nonlinear form, and also the embedding of the structural information itself is more intricate. We suggest that maps of similar type could be explored for different systems and application problems in chemistry.

\section*{Discussion}

Our study demonstrates that ``synthetic'' atomic energies, predicted at large scale by a machine-learning model, are themselves learnable and can be used to study the behaviour of different atomistic ML techniques. 
In the present work, we have compared the ability of GPR, NN, and DKL models to predict properties of chemical environments in the large-data limit, using atomic energies as a proxy for other quantities.

By means of numerical experiments, we showed that network-based models are able to learn useful representations from the original SOAP descriptors, and that these can lead to improved accuracies compared to SOAP-GPR models if the number of training data points is large.
Whereas DKL can substantially outperform stand-alone NNs in some applications,\cite{yuDeepKernelLearning2020} and has begun to be successfully applied to research questions in chemistry, \cite{liu_exploring_2022, sivaraman_coarse-grained_2022} we have found that in the setting of the present work ({\em i.e.}, regression of large amounts of per-atom energy data), DKL models were only slightly more accurate than NNs whilst being significantly more expensive when making predictions.
In comparison, SOAP-based GPR models, while slower and less accurate in the large-data regime, show better generalisation (and thus accuracies) for small amounts of data. This finding is consistent with the marked success of sparse-GPR-based atomistic ML models on datasets of (relatively) modest size. \cite{deringerGaussianProcessRegression2021}

In the present work, we have focused on learning ML atomic energies. 
These values do not directly correspond to any quantum-mechanical observable, and yet empirically they do appear to correlate well with local topological disorder and distortions,\cite{bernsteinQuantifyingChemicalStructure2019a} and they can be used to drive structural exploration. \cite{el-machachiExploringConfigurationalSpace2022a}
Irrespective of their physical interpretation (or absence thereof), we propose that ML atomic energies are a useful regression target for NN models: the compressed representations of structure learned through this task are imbued with deep, and to some extent interpretable, meaning (Fig.\ \ref{fig:embeddings}). 
The fact that synthetic labels are quick to generate, and the networks quick to train, suggests that this is a useful auxiliary or pre-training task to create models with ``knowledge'' about a chemical space.
These network models could then be used to fine-tune on much smaller datasets, using their existing and general chemical knowledge to overcome the relative weakness of NN models in the low-data regime. We have shown initial evidence for this in Fig.\ \ref{fig:pre-training}, and further work is ongoing.

\section*{Appendix: Technical details}
When training all NN models, we employed as means of regularisation: early stopping (by measuring performance on a validation set), dropout, and L2 weight decay. 
To find optimal parameters for the number of hidden layers, layer width, learning rate, batch size, dropout fraction, and weight decay magnitude, we performed a random (Hammersley) search over a broadly defined hyper-parameter space. 
We found that the optimal learning rate in all instances was close to the commonly used $3 \times 10^{-4}$ for the Adam optimiser. 
3 hidden layers, each with $\approx 800$ nodes, gave the most accurate models in the limit of large data, while smaller models performed better for smaller training sets, presumably because the model size is acting as further regularisation to avoid extreme overfitting. 
In all instances, we found high dropout ($p \approx 0.5$) to be a more effective regulariser than weight decay.

For the NN models trained on DFT per-cell energies, we obtained errors on the test and training set amounting to 52.2 and 26.2 meV atom$^{-1}$, respectively, and for the fine-tuned models we obtained 46.0/21.3 meV atom$^{-1}$ on test/train data.
We note that, despite aggressive regularisation during training, this generalisation gap is large -- we attribute this to the small dataset size (1200 labels) used in the fine-tuning experiment. In comparison, the corresponding generalisation gap for the $N = 10^6$ NN model in Fig.\ \ref{fig:learning-curves} is much smaller, {\em viz.}\ 23.7/22.0 meV atom$^{-1}$ for test/train data, respectively.

\section*{Data availability}
The dataset supporting the present work is provided at \url{https://github.com/jla-gardner/carbon-data}. Each trajectory is supplied as a standalone ``extended'' XYZ file, with local energies provided as a per-atom quantity. These files can be read and processed, for example, by the Atomic Simulation Environment (ASE). \cite{larsenAtomicSimulationEnvironment2017}

\begin{acknowledgements}
 We thank J.\ N.\ Foerster, A.\ L.\ Goodwin, and T.\ C.\ Nicholas for useful discussions.
 J.L.A.G. acknowledges a UKRI Linacre - The EPA Cephalosporin Scholarship, support from an EPSRC DTP award (EP/T517811/1), and from the Department of Chemistry, University of Oxford.
 V.L.D. acknowledges a UK Research and Innovation Frontier Research grant [grant number EP/X016188/1] and support from the John Fell OUP Research Fund.
 The authors acknowledge the use of the University of Oxford Advanced Research Computing (ARC) facility in carrying out this work (http://dx.doi.org/10.5281/zenodo.22558).
\end{acknowledgements}

\end{document}